\documentclass[useAMS,usenatbib,usedcolumn]{mn2e}

\def\HI{\ifmmode{\rm HI}\else{H\/{\sc i}}\fi}

\def\lsun{\ifmmode{{\mathrm L}_{\odot}}\else{L$_{\odot}$}\fi}

\def\msun{\ifmmode{{\mathrm M}_{\odot}}\else{M$_{\odot}$}\fi} 
\def\msunpc2{\ifmmode{{\mathrm M}_{\odot} \, {\mathrm{pc}}^{-2}}\else{M$_{\odot} \, {\mathrm {pc}}^{-2}$}\fi}

\def\kms{\ifmmode{{\mathrm{km \, s^{-1}}}}\else{${\mathrm{km \, s^{-1}}}$}\fi}

\def\la{\mathrel{\mathchoice {\vcenter{\offinterlineskip\halign{\hfil
$\displaystyle##$\hfil\cr<\cr\sim\cr}}}
{\vcenter{\offinterlineskip\halign{\hfil$\textstyle##$\hfil\cr
<\cr\sim\cr}}}
{\vcenter{\offinterlineskip\halign{\hfil$\scriptstyle##$\hfil\cr
<\cr\sim\cr}}}
{\vcenter{\offinterlineskip\halign{\hfil$\scriptscriptstyle##$\hfil\cr
<\cr\sim\cr}}}}}
\def\ga{\mathrel{\mathchoice {\vcenter{\offinterlineskip\halign{\hfil
$\displaystyle##$\hfil\cr>\cr\sim\cr}}}
{\vcenter{\offinterlineskip\halign{\hfil$\textstyle##$\hfil\cr
>\cr\sim\cr}}}
{\vcenter{\offinterlineskip\halign{\hfil$\scriptstyle##$\hfil\cr
>\cr\sim\cr}}}
{\vcenter{\offinterlineskip\halign{\hfil$\scriptscriptstyle##$\hfil\cr
>\cr\sim\cr}}}}}

\def\aj{AJ}
\def\apj{ApJ}
\def\apjl{ApJ}
\def\apjs{ApJS}
\def\aap{A\&A}
\def\mnras{MNRAS}
\def\pasp{PASP}

\usepackage[figuresright]{rotating}
\usepackage{lscape}
\usepackage{graphics}
\usepackage{epsfig}
\usepackage{multirow}
\usepackage{bigdelim}
\usepackage{bigstrut}
\usepackage{amsmath}
\usepackage{amssymb}
\usepackage{hyperref}
\usepackage{epstopdf}

\defcitealias{Jaffe2014}{J14}

\title[BUDHIES II]{BUDHIES II:  A phase-space view of HI gas stripping  and star-formation quenching 
in cluster galaxies} 
\author[Y.~Jaff\'e et al.] {Yara L. Jaff\'e$^{1}$\thanks{E-mail: yara.jaffe@astro-udec.cl}, 
Rory  Smith$^1$, Graeme N. Candlish$^{1}$, \and  Bianca M. Poggianti$^2$,  Yun-Kyeong Sheen$^1$,  Marc A.~W. Verheijen$^3$ \\ 
   $^1$Department of Astronomy, Universidad de Concepci\'on, Casilla 160-C, Concepci\'on, Chile\\
   $^2$INAF - Osservatorio Astronomico di Padova, vicolo dell' Osservatorio 5, I-35122 Padova, Italy\\
   $^3$University of Groningen, Kapteyn Astronomical Institute, Landleven 12, 9747 AD, Groningen, The Netherlands }
   \date{Accepted 2015 January 7.  Received 2015 December 22; in original form 2014 September 15}

\begin{document}

\maketitle

\begin{abstract}
We investigate the effect of ram-pressure from the intracluster medium on the stripping of HI gas in galaxies in a massive, relaxed, X-ray bright, galaxy cluster at $z=0.2$ from the Blind Ultra Deep HI Environmental Survey (BUDHIES). 
We use cosmological simulations, and velocity vs. position phase-space diagrams to infer the orbital histories of the cluster galaxies.  In particular, we embed a simple analytical description of ram-pressure stripping in the simulations to identify the regions in phase-space where galaxies are more likely to have been sufficiently stripped of their HI gas to fall below the detection limit of our survey. 
We find a striking agreement between the model predictions and the observed location of HI-detected and non-detected blue (late-type) galaxies in phase-space, strongly implying that ram-pressure plays a key role in the gas removal from galaxies, and that this can happen during their first infall into the cluster.  
However, we also find a significant number of gas-poor, red (early-type) galaxies in the infall region of the cluster that cannot easily be explained with our model of ram-pressure stripping alone.  
We discuss different possible additional mechanisms that could be at play, including the pre-processing of galaxies in their previous environment. 
Our results are strengthened by the distribution of galaxy colours (optical and UV) in phase-space, that suggests that after a (gas-rich) field galaxy falls into the cluster, it will lose its gas via ram-pressure stripping, and as it settles into the cluster, its star formation will decay until it is completely quenched. 
Finally, this work demonstrates the utility of phase-space diagrams to analyze the physical processes driving the evolution of cluster galaxies, in particular HI gas stripping.

\end{abstract}


\section{Introduction}
\label{sec:introduction}

Much work has been done to understand the physical processes driving galaxy evolution.
It has been shown by many observational and theoretical studies that the environment plays an important role driving the evolution of galaxies. \citep[e.g.][]{Dressler1980,Lewis2002,gomez03,Koopmann2004,BoselliGavazzi06,poggianti2006,Moran2007,Bamford2009,Jaffe2011b}.
An unassailable result from observations of galaxies in the local universe is that cluster galaxies are more likely to have an elliptical or S0 morphology than field galaxies, which are typically star-forming spirals \citep{Dressler1980}. 
In addition, it is known that the morphology of galaxies correlates with their stellar populations. In a colour-magnitude diagram, elliptical and S0 (i.e. early-type) galaxies follow a tight correlation \citep[the so-called ``red sequence"; ][]{Baum1959,VS1977}, that indicates they have old stellar populations \citep{BLE,Aragon1993,Jaffe2011}. Spiral and irregular (i.e. late-type) galaxies instead tend to display a wide range of bluer colours, often referred to as the ``blue cloud". The existence of a clear colour bimodality in clusters strongly supports the idea that environment plays an important role in quenching the star formation in galaxies.     
Moreover, observations of distant galaxies have shown a steady rise in the fraction of S0s in clusters, along with a comparable decline in the spiral fraction since $z\sim0.5$, while the fraction of elliptical galaxies remains constant \citep{Fasano2000,Desai2007}. When considering the predicted evolution of large-scale structure from cosmological simulations \citep[e.g.][]{deLucia2012}, this result suggests that spiral galaxies transform into S0s as they transition from the field into clusters. The physical mechanism(s) driving such a transformation is however still under debate.

A particularly important ingredient for understanding galaxy evolution is the atomic gas. 
Neutral hydrogen (HI) not only provides a reservoir from which $H_{2}$ forms, fueling star formation, but is also a sensitive tracer of different environmental processes, such as ram-pressure stripping and tidal interactions.
\citet{GunnGott1972} first predicted (analytically) that ram-pressure can be effective in removing the galaxies' interstellar medium (ISM) as they pass through a dense intracluster medium (ICM), thereby quenching their star formation. 
 Observations   \citep[e.g.][]{Cayatte1990,BravoAlfaro2000,BravoAlfaro2001,PoggiantiVG2001,Kenney2004,crowl05a,Chung2007,Chung2009,Abramson2011,Scott2010,Scott2012,Gavazzi2013} have shown that the HI gas in galaxies is indeed disturbed and eventually truncated and exhausted in clusters.
Recent simulations \citep[see][for a review]{Roediger2009} have further supported the idea that this happens  via ram-pressure stripping  and gravitational interactions \citep[e.g. ][]{Vollmer2001,Vollmer2003,TonnesenBryan2009,Kapferer2009}. 
It is important to note that ram-pressure is expected to be more effective at the centre of massive clusters, because of the higher densities of the ICM, and the high velocities of the galaxies. 
In fact ram-pressure stripping models indicate that HI can be fully depleted after a transit through the ICM's core \citep{Roediger2007}. 
Moreover, the galaxies in massive clusters also suffer fast encounters with other cluster members. 
The accumulation of such encounters, known as harassment \citep[][]{Moore1999}, can cause disk thickening and gas fuelling of the central region. 
Harassment can strip HI from galaxies \citep{DucBournaud2008} and is most effective in low-surface-brightness galaxies, and towards the core of galaxy clusters \citep[][]{Moore1999}. 
Finally, it has also been suggested that galaxies can be pre-processed in less massive groups before they fall into clusters \citep[e.g.][]{ZabludoffMulchaey1998}. This idea has recently received observational \citep{Balogh1999,Lewis2002,VerdesMontenegro2001,Jaffe2012} and theoretical support \citep{McGee2009,deLucia2012}.  
In order to acquire a complete understanding of the physical processes affecting galaxies through their lifetimes, it is thus necessary to couple their observed properties to the cosmic  environments which they have inhabited throughout their lifetimes. 

Due to technological limitations, studies of the HI content of galaxies have so far mostly been carried out in the local universe ($\lesssim0.08$). 
To investigate the relation between gas reservoirs and star formation, as well as its evolution, we carried out a Blind Ultra Deep HI Environmental Survey \citep[BUDHIES][]{Verheijen2007,Jaffe2013}, that focuses on two pencil-beam pointings containing 2 galaxy clusters at $z \simeq 0.2$, and the large-scale structure around them. The clusters,  Abell 2192 and Abell 963 (A2192\_1 and A963\_1 from now on\footnote{As in \citet{Jaffe2013}, we adopt the names  ``A963\_1" and  ``A2192\_1"  to distinguish these clusters from the other structures we found along the same line-of-sight, such as galaxy groups and cosmic sheets.}) are very different from each other.

In \citet{Jaffe2012} we studied  A2192\_1, and found that this intermediate-mass cluster has a significant amount of substructure \citep[see also][]{Jaffe2013}. The incidence of HI-detections significantly correlates with environment: at large clustercentric radii ($\gtrsim $2-3 virial radii), where the infalling substructures were found,  many galaxies are detected in HI, while at the core of the forming cluster, none of the galaxies are HI-detected. We found that this effect starts to become significant in low-mass groups that pre-process the galaxies before they enter the cluster, so that  by the time the group galaxies fall into the cluster they may already be HI deficient or even devoid of HI.  

A963\_1, on the other hand, is a massive cluster with significant X-ray emission centred on the brightest cluster galaxy (BCG). This cluster is thus ideal to study the interaction between galaxies and the ICM.  In this paper, we focus on A963\_1 to study the effect of ram-pressure in the stripping of HI gas from cluster galaxies, and the later quenching of their star formation. 

Recent works by \citet{Mahajan2011,Oman2013,HernandezFernandez2014,Muzzin2014,Taranu2014} have highlighted the usefulness of  phase-space diagnostics to infer the assembly histories of clusters. 
In simulations it is possible to trace back the orbital histories of galaxies in clusters, and associate different populations (e.g. infalling, backsplash, or galaxies in the virialized part of the cluster) to different phase-space locations. 
Simulations can provide a 6-dimensional phase-space (all 3 components of both the position and velocity vectors: $x$,$y$,$z$,$v_x$,$v_y$,$v_z$ respectively). Observations of distant galaxies, however, only give 3 projected parameters: 2 coordinates, and the line-of-sight velocity ($\alpha$, $\delta$, and $v_{los}$ respectively). For simplicity we refer to the simulated phase-space as ``3D", and to the observed one as ``projected". 
Despite the loss of information, it is still possible to infer the assembly history of cluster galaxies from their projected phase-space \citep[PPS; see e.g.][]{Oman2013,HernandezFernandez2014}.  The uncertainties introduced by the projection effects can be minimized by utilizing large (mass-limited) samples of galaxies in regular clusters of similar mass and redshift. 
Moreover, this approach certainly provides a more complete view of environment compared to the standard 1D view (galaxy properties as a function of radial clustercentric distance only). 

In this paper, we investigate the environmentally driven evolution of galaxies in A963\_1, a massive cluster at $z=0.2$, making use of the unique BUDHIES dataset. We exploit the HI-data and ancillary data to map the gas stripping and later suppression of the star formation activity of galaxies in phase-space. 
To interpret our data, we compare it to a simple ram-pressure model built into a cosmological simulations, that predicts the regions in phase-space where HI gas is expected to be stripped from the galaxies in our cluster. 

The paper is organized as follows: 
In Section~\ref{sec:data} we summarize the data collected for BUDHIES (Sec.~\ref{subsec:budhies}), the main characteristics of the cluster (Sec.~\ref{subsec:cluster}), and the properties of its member galaxies (Sec.~\ref{subsec:obs_props}). 
In Section~\ref{sec:obs_pps} we show how the cluster galaxies are distributed in PPS, highlighting the location of the HI-detections. In Section~\ref{sec:rps} we create a simple ram-pressure stripping model, which we embed in the cosmological simulations presented in Section~\ref{sec:sims}. These are, in turn, used  to derive the assembly histories of the observed clusters and predict the final location of stripped galaxies in PPS. In Section~\ref{sec:bigpic} we combine our observed and simulated results and discuss the effect of ram-pressure stripping, and compare our results with previous work. We also use galaxy colours in the PPS analysis as tracers of post-stripping evolution in Section~\ref{sec:Poststrip}, to better understand the late stages of star-formation quenching as a function of the orbital histories of the cluster galaxies.  We summarize our findings and draw conclusions in Section~\ref{sec:conclu}. 

Throughout this paper we assume a concordance $\Lambda$CDM cosmology with $\Omega_{\rm M}=$0.3, $\Omega_{\Lambda}=$0.7, and H$_{0}=$70 km s$^{-1}$ Mpc$^{-1}$.


\section{Observations}
\label{sec:data}

\subsection{BUDHIES data}
\label{subsec:budhies}

BUDHIES is an ultra-deep HI survey of galaxies in and around two clusters at $0.16\leq z\leq0.22$. 
The survey has an effective volume depth of 328 Mpc and a coverage on the sky of $\sim$12$\times$12 Mpc for each cluster, which allows us to study the large-scale structure around the clusters.  
At the core of our study are the ultra-deep HI observations, carried out at the Westerbork Synthesis Radio telescope \citep[WSRT][]{Verheijen2007,Verheijen2010} from which we have detected 127 galaxies in A963\_1 and 36 in A2192\_1  with  HI masses $\gtrsim 2 \times 10^9 M_{\odot}$. 
In addition to the HI data, we have obtained $B$- and $R$-band imaging with the Isaac Newton Telescope (INT), NUV and FUV imaging with GALEX (Fernandez et al. in preparation), Spitzer imaging, 
and near-infrared imaging with UKIRT. 
SDSS photometry is available for the galaxies in the field, and in a limited number of cases, SDSS spectra are also available. 
Finally, we have carried out a spectroscopic campaign with WIYN and the AF2 on the William Herschel Telescope (WHT), from which we have assessed cluster memberships and characterized the environment in the surveyed volumes \citep[][]{Jaffe2013}.

\subsection{A963\_1}
\label{subsec:cluster}

With a dynamical mass of $1.1 \times 10^{15} h^{-1} M_{\odot}$ ($\sigma_{cl}= 993$ km s$^{-1}$), A963\_1 is the most massive cluster in BUDHIES.   
It is found at $z=0.2039$ and its $R_{200}$ was estimated to be 1.55~$h^{-1}$Mpc \citep{Jaffe2013}. 
The cluster's total X-ray luminosity is $L_X \backsimeq 3.4 \pm 1 \times 10^{44} \text{erg} s^{-1} h^{-2}$ \citep[][]{Allen2003}. Its regular and centrally concentrated  X-ray emission is centered on a central cD galaxy \citep[see Fig. 6 of ][]{Smith2005}, suggesting A963\_1 is a relaxed cluster. 
However, in \citet{Jaffe2013} we performed a  Dressler-Shectman test  \citep{DresslerShectman1988} on the spectroscopic members of A963\_1 out to larger radii, and  concluded  that the ``relaxed" part of the cluster is at its core, while there is a mild presence of substructure in the outskirts. In addition,  small X-ray substructures have been observed at $r>R_{200}$ \citep[see Fig. B.6. in][see also Haines et~al. in prep. for a full description of the X-ray substructures]{Zhang2007}. 
In this paper, we focus on the global properties of A963\_1, ignoring small inhomogeneities in the ICM. 
This is justified because the vast majority of the X-ray flux is centrally concentrated and very regular. The substructure in the cluster outskirts does not dominate.    
We leave the detailed study of the HI properties of galaxies as a function of X-ray structure for a future paper. 

\subsection{The cluster galaxies}
\label{subsec:obs_props}

In the following, we present derived properties of the member galaxies of A963\_1. 

\subsubsection{Stellar masses}
To compute stellar masses ($M_{\star}$) for all the spectroscopically confirmed cluster members, we first converted the 5-band SDSS photometry available to absolute magnitudes using our spectroscopic redshifts, and the interpolation utility InterRest \citep{Taylor2009}. We then computed the stellar masses following the prescription by \citet{Zibetti2009}, with a Kroupa initial mass function. 

\begin{figure}
\centering
\includegraphics[scale=0.47]{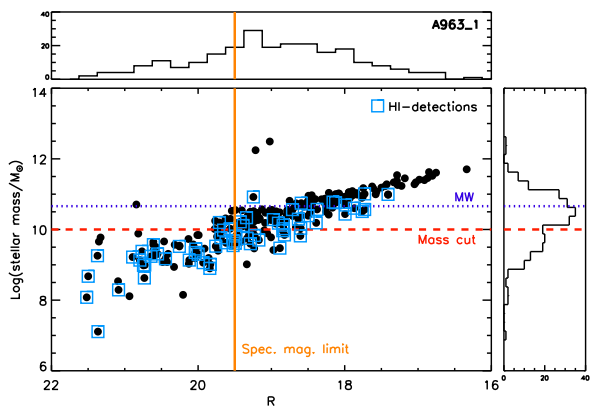}
\caption{ R-band magnitude versus $\log(M_{\star})$ for the cluster members (black circles). 
The HI-detections are distinguished with open blue squares. At the top and right-hand side of the plot, histograms of the $R$-band magnitude and $\log(M_{\star})$ are also shown. The dashed red line indicates the mass cut adopted  in the analysis of Section~\ref{sec:bigpic} (see also Section~\ref{sec:obs_pps}), and the dotted blue line is the mass of the Milky Way-like galaxy used as a prototype to explore the effects of ram-pressure stripping in the cluster (c.f. Section~\ref{sec:rps}). The magnitude limit adopted in the spectroscopic campaign is indicated by the vertical orange line.  
 \label{Ms_a963}}
\end{figure}

The distribution of $M_{\star}$ for the cluster members is shown in Figure~\ref{Ms_a963}. 
In the figure,  $\log(M_{\star})$ is plotted against $R$ magnitude.  HI-detected galaxies are marked with a blue square and occupy the region of lower stellar masses and fainter R magnitudes. The horizontal dashed red line indicates the stellar mass cut ($M_{\star}=10^{10} M_{\odot}$) used to define a mass-limited sample, adopted in some of the analysis presented in this paper. This mass cut roughly matches the magnitude cut where our spectroscopic campaign is most complete (dashed orange vertical line,  see \citet{Jaffe2013} for details on the target selection).

\begin{figure*}
\centering
\includegraphics[scale=0.46]{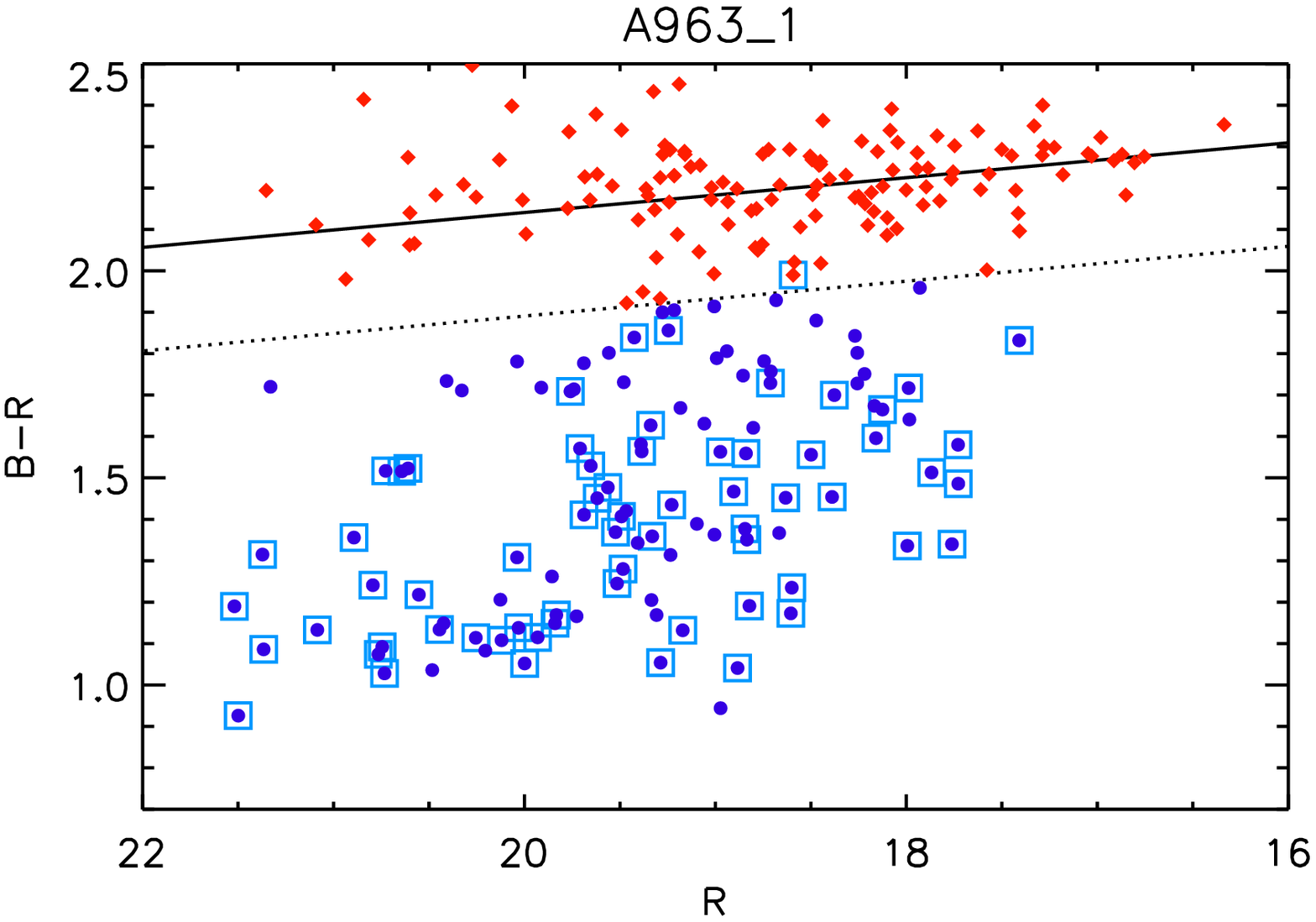}
\includegraphics[scale=0.46]{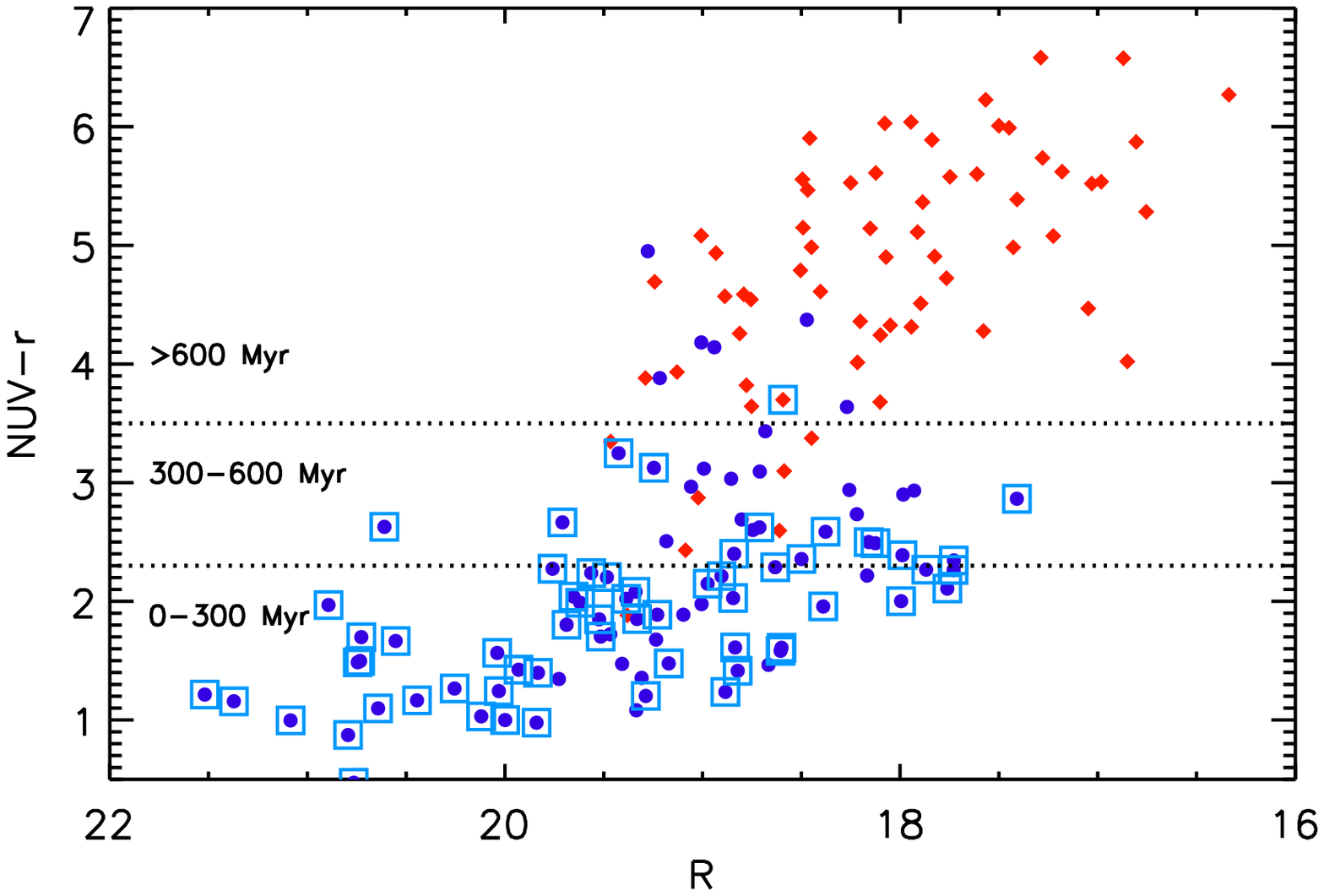}
\caption{\textit{Left:} B-R vs R diagram of galaxies in in A963\_1.  
The red sequence fit is shown by the solid line. Red-sequence galaxies (red diamonds) are defined to be at most $\pm0.25$mag bluer than the red sequence fit (dotted line). 
The blue-cloud galaxies (blue circles), are the ones that lie below the dotted line. \textit{Right:} NUV-optical colour-magnitude diagram showing the red-sequence and blue-cloud galaxies defined in the left panel, and dotted lines separating regions (colour ranges) that correspond to different stellar-population ages, as labeled (see text for description). 
In both panels HI detections are highlighted with open blue squares. 
\label{CMDs}}
\end{figure*}

\subsubsection{Colours}
\label{subsec:cols}

To distinguish different galaxy populations, we constructed colour-magnitude diagrams (CMDs) using the UV and optical data. 
In the left panel of Figure~\ref{CMDs}, we show the  $B-R$ vs. $R$  CMD for A963\_1.  
We fitted the red sequence 
using the reddest ($B-R>1.9$) early-type galaxies (i.e. elliptical and S0s). 
The best fit ($B-R = -0.042 \times R + 2.98$) is shown by the solid black line in the plot. We further define the blue cloud  to contain all galaxies bluer than 0.25mag below the red sequence (blue filled symbols below dashed line). 
We note that the morphological classification used to define the red sequence was the product of visual inspection of the INT $B$ and $R$ images. Several co-authors assigned morphologies to the galaxies and these were later combined. The quality of the images, however, was too poor to robustly assign morphologies to all the cluster galaxies.  In fact we could only determine a rough morphological type for a small sample of galaxies. To avoid introducing scatter to our analysis, or having to reduce the sample significantly, in this paper we only used the morphological information to fit the red sequence. In the analysis however, we use colours instead to broadly separate star-forming, late-type galaxies (blue) from those with quenched star formation and early-type morphology (red).

The colour classification shown in the left panel of Figure~\ref{CMDs} provides an indication of the ages of the stellar populations in these galaxies. The majority of red-sequence galaxies, for instance, are expected to have stopped forming stars, while blue galaxies are still hosting star formation and thus contain a significant fraction of young stellar populations. 
In the Figure~\ref{CMDs} we have also highlighted the location of HI-detected galaxies.  
As expected, these populate the blue cloud.  

Table~\ref{sample_table} lists the sizes of the total ``blue'' and ``red'' galaxy samples, as well as the number of galaxies in the mass-limited and HI-detected samples.  

The right panel of Figure~\ref{CMDs} shows the UV-optical colour-magnitude diagram. 
UV is very sensitive to the ages of young stellar populations. It is therefore helpful to utilize NUV-optical colours as a tool to roughly infer the ages of the young stellar component in the galaxies.  We do this by inspecting the NUV-r colour evolution in a single-stellar population model of solar metallicity from \citet{bc03}\footnote{We note that our results do not change if we use the models from Starburst99 \citep{Leitherer1999}}. 
We adopt the same x-axis range and symbol colour-code as in the left panel to easily locate the red-sequence and blue-cloud galaxies previously defined in optical bands. 
As expected, the galaxies spread over a larger range in colour in the UV-optical colour-magnitude relations. 
We choose three NUV-r colour bins that correspond to the ages of a single stellar population of $<$ 300 Myr; 300$-$600 Myr; $>$ 600 Myr, at the redshift of our target cluster. 
These bins are identified by the dotted lines in the right panel  of Figure~\ref{CMDs}. We emphasize that these lines should only be treated as good reference values, and not exact ages of the stellar populations. 
We note that we did not consider internal extinction in our computation, but we use these colour cuts only as a lower limit for a given age of the young single stellar population.
We will use these colour bins in the analysis of Section~\ref{sec:bigpic}.

\begin{table}
\begin{tabular}{llll}
\hline
 				 & No. gals. &  No. massive gals. & No. HI-detec.\\
\hline
Red 	&  128 & 109 & 1			\\
Blue 	& 118 & 49 & 65			\\
\hline
\end{tabular}
\caption{The galaxy subsamples used in this paper, split by their optical colours, as defined in the left panel of Figure~\ref{CMDs}. The colums are: sample name (``Red" and ``Blue", corresponding to the red sequence and blue cloud), number of galaxies, number of galaxies in the mass-limited sample ($M_{\star}>10^{10}M_{\odot}$), total number of HI-detected galaxies.  }
\label{sample_table}
\end{table}

\section{A phase-space view of A963\_1}
\label{sec:obs_pps}

\begin{figure*}
\centering
\includegraphics[scale=0.75]{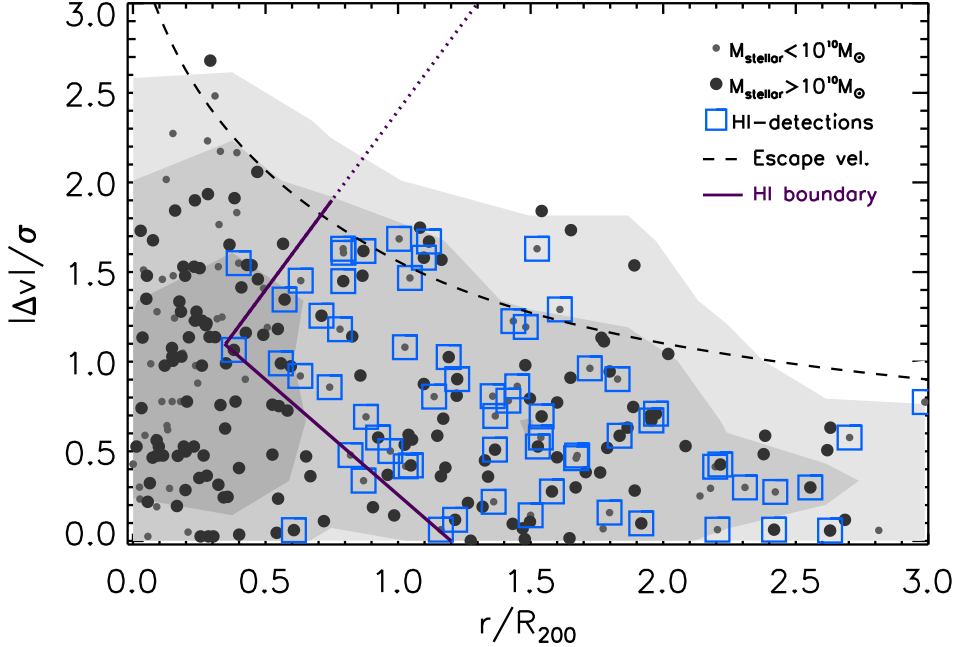}
\caption{The observed PPS for all galaxies in A963\_1. 
Galaxies in the mass-limited sample ($M_{\star}>10^{10}M_{\odot}$) are represented with darker and larger filled symbols, HI-detected galaxies are enclosed by a blue open square, and  the grey countours follow the number density of the galaxies. The dashed black line corresponds to the escape velocity in a NFW halo (see text for details).  
The solid purple line delimits the region (right of the line) containing the vast majority of the HI-detections in the cluster. 
\label{PPS_obs1}}
\end{figure*}

To  investigate the dynamical state of the cluster under study and infer (statistically) the orbital histories of the different galaxy populations, we have constructed line-of-sight velocity vs. radial position phase-space diagrams for the spectroscopically-confirmed members, as shown in  Figure~\ref{PPS_obs1}. 
The x-axis of the plot shows the projected distance from the cluster centre, $r$, normalized by $R_{200}$, while the y-axis displays the peculiar line-of-sight velocity of each galaxy with respect to the cluster recessional velocity ($\Delta v$), normalized by the velocity dispersion of the cluster. In particular, we plot the absolute value, taking advantage of the symmetry in the velocity distribution of the cluster. This is: 
\begin{equation}
\frac{|\Delta v|}{\sigma} = \frac{|c(z-z_{cl})|}{(1+z_{cl})\sigma} 
\end{equation}
where $z$ is the redshift of a given galaxy, $z_{cl}$ the redshift of the cluster, and $c$ the speed of light.

One can think of the position of galaxies in PPS as lower limits to their real 3D radius and velocity values. In other words, a galaxy cannot be at a smaller radius as it appears in projection. Likewise, it cannot have a smaller velocity difference than its line-of-sight velocity difference.

The grey contours in Figure~\ref{PPS_obs1} show the galaxy number density distribution in phase-space.  
This plot represents a snapshot (at the observed redshift) of the cluster assembly process, where infalling galaxies come towards the cluster at large radii, gaining velocity as they reach the cluster centre. They will then settle into the cluster, partly by means of dynamical friction (which is especially important for high mass galaxies), creating the dense  ``wedge" in PPS, as we will see in Section~\ref{sec:sims}. The overall distribution of cluster galaxies in PPS thus arises from ``virialization", and it is constrained by the radial dependence of the escape velocity in PPS. This is shown by the dashed black line, that corresponds to the escape velocity in a \citet*[][NFW]{nfw96} halo, projected to the line-of-sight. This is:
\begin{equation}
v_{esc} = \left\{
        \begin{array}{ll}
            \sqrt{\frac{2GM_{200}}{3R_{200} }K} & \quad r<R_{200} \\
            \sqrt{\frac{2GM_{200}}{3 R_{200} s}} & \quad {\rm otherwise}
        \end{array}
    \right.
\end{equation}

Where 

\begin{equation}
K=g_c\left(\frac{\ln(1+cs)}{s} -\ln(1+c)\right)+1
\end{equation}

\begin{equation}
s=\frac{r_{_{3D}}}{R_{200}}\sim \frac{\pi}{2} \frac{r}{R_{200}}
\end{equation}

\begin{equation}
g_c=\left[\ln(1+c)-\frac{c}{(1+c)}\right]^{-1}
\end{equation}

 and $c$ is the concentration parameter, which we fixed at $c=6$ (the exact value of $c$ does not change the results significantly). Note that the  3D radius scales with the projected $r$, with scatter, as: $r = \cos(\varphi) r_{_{3D}}$, where $\varphi$ is the projection angle, that can vary between 0 and $\pi/2$. Therefore, on average, $r \sim \frac{2}{\pi} r_{_{3D}}$.  We also projected the 3D escape velocity ($v_{esc}$) to a line of sight escape velocity ($v$) using the average conversion factor. Inevitably, the projections (i.e. in both $v$ and $r$) lead to increased  scatter in PPS. 
In the figure, we have further normalized $v$ to the velocity dispersion of A963\_1. Galaxies inside the dashed black line in Figure~\ref{PPS_obs1} are expected to be gravitationally bound to the cluster. 

The  ``virialized" region of the cluster is roughly located at $r \lesssim R_{200}$ and $\Delta v < 1.5 \sigma$ \citep[see e.g.][]{Mahajan2011}, although the shape of the virialized region is best approximated by a triangle, rather than a square (as will be shown in Figures~\ref{orbits},~\ref{simsplot},~\ref{PPS_obs}~and~\ref{PPS_obs_UV}). 
70\% of the galaxies inside this region have been inside the cluster ($r_{3D}<R_{200}$) for more than 4 Gyrs, which corresponds to the crossing time\footnote{A ``crossing time'' refers to the time that an infalling galaxy takes to cross from $r=R_{200}$ to the other side of the cluster.} of our cluster. In other words, 
the virialized region contains galaxies that have most likely experienced many pericentric passages.

In the PPS of Figure~\ref{PPS_obs1}, we have  distinguished low-mass galaxies from more massive ones with symbol sizes, and we can see that low- and high-mass galaxies are roughly evenly distributed across PPS in the cluster (although there are regions notably dominated by massive galaxies, such as $0.3 \lesssim r/R_{200} \lesssim 1$ and $|\Delta v|/\sigma \lesssim0.7$). This is not necessarily a real effect but perhaps reflects the spectroscopic incompleteness of the low-mass galaxy sample.  
Although we always plot all galaxies for reference, it is important to be aware that a significant assessment of the incidence of different galaxy populations can only be done in the mass-limited (or spectroscopically complete) sample. For this reason we mostly focus on the high-mass galaxies, plotted with bigger symbols in our analysis. 

In Figure~\ref{PPS_obs1} we also highlight, for the moment,  HI-detected galaxies (blue squares), as these observations are the core of the survey. Later, in Figures~\ref{PPS_obs} and~\ref{PPS_obs_UV}, we will see how other galaxy properties are distributed in PPS. 
It is clear that the HI-detected galaxies are located outside the cluster virialized region  and that there is also a notable lack of HI-detections outside the virialized region at high velocities and small radii. We indicate with the solid purple line in Figure~\ref{PPS_obs1} the boundary between the ``HI-rich" region (right of the line) and the region devoid of HI-detections (or ``HI-poor"; left of the line). Note that this line was drawn by eye, as reference. The cause of this segregation between the HI-detected and non-detected galaxies will be thoroughly investigated in the following sections. 

We emphasize that all HI-detected galaxies were plotted in this diagram, no HI mass cut was assumed. However, the effect of the primary beam attenuation of the WSRT translates into a variation of the HI-mass limit with distance from the pointing centre. 
We will describe these effects in Verheijen et al. (in preparation). 
It is remarkable that we see strong decline of the incidence of HI-detected galaxies towards the cluster centre (see also Figure~\ref{rad_trends}) despite this effect, which makes HI harder to detect at larger radii. 

\section{The impact of ram pressure stripping in PPS}
\subsection{Ram-pressure stripping model} 
\label{sec:rps}

To explain the segregation of HI-detected galaxies in PPS (Section~\ref{sec:obs_pps}), we explore the effect of ram-pressure stripping in A963\_1. 
We do this by   creating a simple model, based on the analytic description provided in \citet{GunnGott1972}. 
In particular, following \citet{HernandezFernandez2014}, we compute the ``intensity" ($\eta$) of ram-pressure ($P_{ram}$) exerted by the intra-cluster medium on an infalling galaxy, as the ratio of such pressure over the anchoring self-gravity provided by the galaxy ($\Pi_{gal}$). That is,  $\eta=P_{ram}/\Pi_{gal}$.

We model the ICM gas density profile of the cluster with the standard $\beta$-model \citep{Cavaliere1976}, as:
\begin{equation}
\rho_{_{ICM}}(r_{_{3D}})=\rho_0 \left[1+\left(\frac{r_{_{3D}}}{R_c}\right)^2\right]^{-3\beta/2}
\end{equation}
where, as before, the  3D radius scales with the projected $r$, with scatter, as: 
$r \sim \frac{2}{\pi} r_{_{3D}}$.

Now, the ram-pressure is defined as: 
\begin{equation}
P_{ram}=\rho_{_{ICM}} v_{gal}^2
\end{equation}

Again, $v_{gal}$ is the velocity of the galaxy in 3D, which we also project to the line-of-sight velocity, $v$. 

We then model the galaxy's anchoring pressure, $\Pi_{gal}$, as the following: 

\begin{equation}
\Pi_{gal}=2 \pi G \Sigma_{\rm s}\Sigma_{\rm g} 
\end{equation}

Where $\Sigma_{\rm s}$ and $\Sigma_{\rm g}$ are the density profiles of the stellar and gaseous disks respectively. We assume an exponential profile for the disk:

\begin{equation}
\Sigma=\Sigma_0 e^{-r'/R_d}
\label{sigma0}
\end{equation}

where $r'$ is the radial distance from the centre of the galaxy,   $R_d$ is the disk scalength, and $\Sigma_0$ the central surface density of the galaxy:

\begin{equation}
\Sigma_0=\frac{M_d}{2 \pi R_d^2}
\label{eq_gal_sd}
\end{equation}

where $M_d$ is the mass in the disk.

We adopt the $\beta$-model parameters from \citet[][]{Rizza1998}   
to compute $P_{ram}$ for A963\_1. These values are summarized in  Table~\ref{Cluster_param}.

\begin{table}
\begin{tabular}{ll}
\hline
A963\_1's ICM model & \\
\hline
$\beta$ 		& 0.5 										\\
$R_c$ 		& 65~kpc									\\
$\rho_0$		& 14.7$\times10^{-3}$~cm$^{-3}$			\\
\hline
\end{tabular}
\caption{Parameters of the $\beta$-model for A963\_1, used to compute the intensity of ram-pressure as a function of distance from the cluster centre. The values were taken from  \citet{Rizza1998}. }
\label{Cluster_param}
\end{table}

Finally, we construct a disk galaxy model  
to test the effect on ram-pressure on it. The values used to model the galaxy's anchoring force (listed in  Table~\ref{MW_values}) are a reasonable representation of a Milky Way (MW)-like disk galaxy. Throughout the rest of this paper we thus refer to the model as a ``MW" model.  Note that the stellar mass of this test galaxy corresponds to the most typical mass in our mass-limited sample (see dotted blue lines in Figure~\ref{Ms_a963}). Using these values we obtain $\Pi_{gal}=1.24\times 10^{-11}$N m$^{-2}$ at the centre of the galaxy (i.e. at $r'=0$, where the restoring force is the strongest). Note that this value is smaller than the one used by \citet{HernandezFernandez2014}. 

\begin{table}
\begin{tabular}{lll}
\hline
 	&  value 					& referece \\
\hline
$M_{\rm d,stellar}$ 	& 	4.6$\times 10^{10}$M$_{\odot}$	& 	\citet{Bovy2013}	\\
$M_{\rm d,gas}$ 		& 	$\sim$0.06$\times M_{\rm d,stellar}$		&  \citet{Gavazzi2008} 	\\
$R_{d, \rm stars}$	& 	2.15 kpc							& 	\citet{Bovy2013} \\
$R_{d, \rm gas}$		& 	1.7	$\times 	R_{d, \rm stars}$	& \citet{Cayatte1994}			\\
\hline
\end{tabular}
\caption{Values used to compute the restoring force in the disk of a MW-type galaxy.}
\label{MW_values}
\end{table}

We can now predict the region in PPS where ram presure is expected to strip the gas off the MW-like galaxy in A963\_1, i.e. the case where $\eta=1$. Jumping ahead to the bottom panel of Figure~\ref{orbits} (or Figures~\ref{simsplot},~\ref{PPS_obs},~and~\ref{PPS_obs_UV}), we can see the case of complete stripping ($r^{\prime}=0$) plotted as a grey dashed line in the PPSs of Figures~\ref{orbits},\ref{simsplot},~\ref{PPS_obs}~and~\ref{PPS_obs_UV}. 
Galaxies infalling into the cluster (from the right-hand side of the line towards the cluster centre), will eventually cross this line, losing their gas reservoirs. 
This line thus constitutes a good measure of the presence of HI in PPS. However, because our HI observations have a detection limit of $\sim 2 \times 10^9$M$_{\odot}$, there will be some galaxies that will fail to be detected before the gas disk is truncated all the way down to $r^{\prime}=0$. In other words, we expect to stop seeing HI-detections at larger $r/R_{200}$ than the grey dashed line predicts. 
To account for this effect, we have re-computed $\Pi_{gal}$ at the truncation radius at which a Milky Way galaxy would fall below BUDHIES' HI detection limit. 
The truncation radius, $r_t>0$, is computed from the remaining gas mass in the galaxy, $f=\frac{2\times 10^9 \rm M_{\odot}}{M_{\rm gas}}$, following: 

\begin{equation}
f=1+\left[e^{-r_t/R_d}\left(\frac{-r_t}{R_d} -1\right)\right] \label{eq_tr}
\end{equation}

Solving for $r_t$ we obtain a value of 6.1 kpc, which yields the new $\Pi_{gal}$ at $r'=r_t$, shown by the dashed green line in Figure~\ref{orbits} (and Figures~\ref{simsplot},~\ref{PPS_obs}~and~\ref{PPS_obs_UV}). Galaxies to the left of this line are thus expected to have been stripped enough to fall below our detection limit. We call this region the ram-pressure stripping cone. 

For future reference and application to other systems, in Appendix~\ref{ap:rpmodel} we have explored how much the cluster and galaxy model affect the mapping of the effect of ram-pressure stripping in PPS.  Interestingly, we find that the effectiveness of ram-pressure in PPS does not vary much with galaxy mass (except for the most massive galaxies). We will come back to this in Section~\ref{sec:bigpic}

We also note there are several caveats to our ram-pressure stripping model, that need to be considered for the analysis. 
 To start, our model depends on at least 6  variables (3 spatial and 3 in velocity), and we are probing its effects in two dimensions. 
Some scatter in PPS due to projection effects (in both radius and velocity) is thus expected. 
We also note that, despite the simplicity of Gunn \& Gott's prescription, their model has been validated by many simulations. Although it is well-known that their formula is best at predicting the stripping radius in the case of face-on stripping, other inclination angles (with respect to the infall trajectory) can lead to more complex behaviour \citep[e.g.][]{Abadi1999,Vollmer2001}. However, the amount of stripping is only significantly altered for nearly edge-on galaxies \citep{Roediger2006}. 
Finally, observations and simulations of galaxies in clusters have revealed that it is possible to have gas stripping at larger radii than what \citet{GunnGott1972} predict \citep[e.g.][]{Kenney2004,Tonnesen2007}, due to the presence of non-smooth substructure in the ICM, which translates into scatter in ram-pressure strength at fixed radius. As mentioned in Section~\ref{subsec:cluster},  we will combine the BUDHIES data with deep X-ray observations to thoroughly investigate this scenario in a future paper. 

Despite all the aforementioned caveats, in Section~\ref{sec:bigpic} we will show how the distribution of HI-detected galaxies in PPS is in exceptionally good agreement with the prediction from our model.

\subsection{Cosmological simulations}
\label{sec:sims}

\begin{figure}
\centering
\includegraphics[scale=0.48]{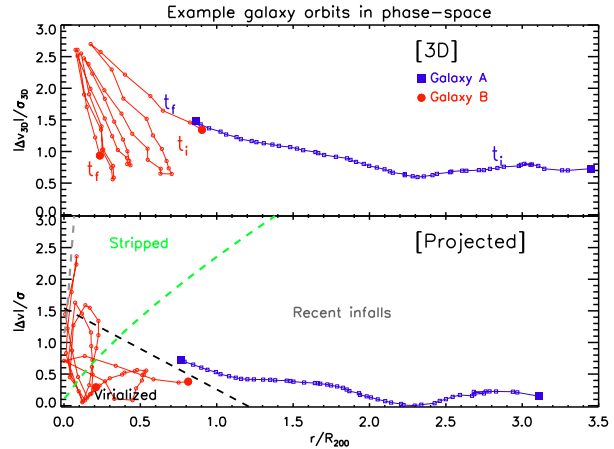}
\caption{The orbit tracks in 3D (top) and projected (bottom) phase-space of two massive galaxies randomly extracted from the simulations. The points correspond to different time-steps in the simulation. The first point in each case is labeled with $t_i$, which corresponds to 7.942 Gyr before the final snapshot, $t_f$. In the bottom panel,  
the dashed grey line indicates the area in PPS (left of the line) where a MW galaxy is expected to be completely stripped as it falls into the cluster. The dashed green line  instead delimits the area  where the model galaxies are stripped enough to fall out of the detection limit of our survey. The dashed black line indicates the region in PPS where galaxies are most likely to have been in the cluster for more than a pericentric passage (i.e. the ``virialized" region). The region to the right of all dashed lines thus contains galaxies that have most likely recently joined the cluster.  
Clearly, at $t_i$ Galaxy A was falling in to the cluster for the first time, while Galaxy B had aready crossed $R_{200}$.  
As time passes, galaxy B oscillates in position and velocity drawing a ``wedge" in phase-space (inside the ``virialized" region). This is what will happen to galaxy A after enough time has passed. \label{orbits}}
\end{figure}

The recipe described in Section~\ref{sec:rps} is a good measure of the velocity and radial dependence of stripping for galaxies on their first infall. However, in PPS we have a mixture of galaxies that are infalling, and others that have been in the cluster for several pericentric passages. To predict the location in PPS where we expect to find stripped galaxies (virialized or infalling), we apply our simple ram-pressure model to 
dark matter halos orbiting in the potential of a cluster formed in a cosmological model. 
The advantage of doing this is that one can trace back the galaxies to study how they build up the cluster (and thus PPS).

\begin{figure*}
\centering
\includegraphics[scale=0.64]{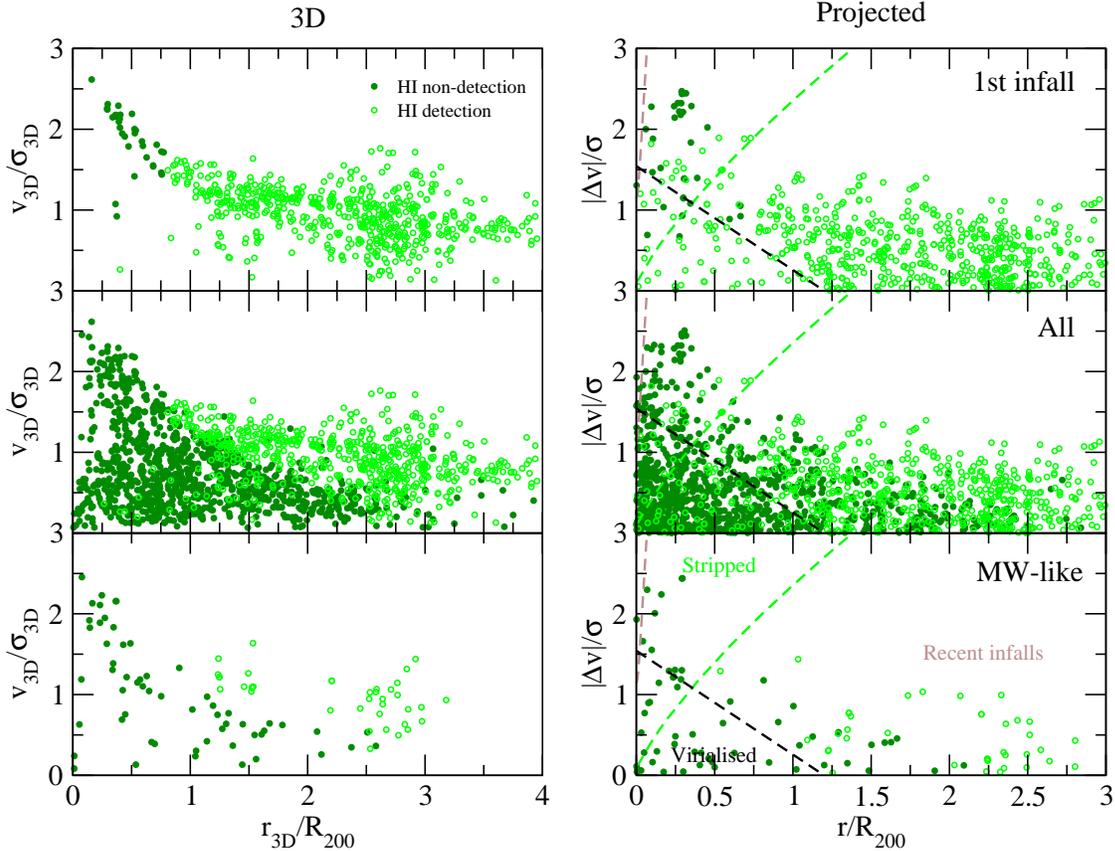}
\caption{The 3D (left) and projected (right) phase-space of the simulated cluster galaxies. The top panels show galaxies that are infalling into the cluster for the first time. They clearly occupy the region in phase-space close to the cluster ``envelope" and far away from its core. The middle panels show all (virialized and infalling) galaxies, which occupy the entire trumped-shaped area of phase-space. At the bottom, we show only massive galaxies (i.e. with masses similar to that of our MW prototype). In the right-hand panels  we further indicate some regions of interest: the ram-pressure stripping cone where galaxies will fall below our HI detection limit (left of the green dashed line), the cone of total HI stripping (left of grey dashed line), and the region in PPS where most galaxies are expected to have been in the cluster for more than a pericentric passage, i.e. the ``virialized" region. \label{simsplot}}
\end{figure*}

We use the high resolution cosmological N-body simulations, fully described in \citet{Warnick2006,Warnick2008}, which were run with the adaptive mesh refinement code MLAPM \citep{Knebe2001}. These simulations used approximately one million dark matter particles, each with a mass of $1.6\times10^8 h^{-1} M_{\odot}$, while the spatial force resolution was $\sim2 h^{-1}$~kpc. The halos were found using AHF, a parallelized version of the MHF algorithm of \citet{Gill2004}. The minimum particle number allowed for a halo was set at 20, leading to a minimum resolved halo mass of $\sim3\times10^9 M_{\odot}$.

To emulate A963\_1 as much as possible, we chose the cluster ``C1" from Table 1 of \citet{Warnick2006}, as it is the most massive cluster in these simulations. The virial mass and radius for that cluster are: $M_{vir}$=$2.9\times10^{14} M_{\odot}$, and $R_{vir}=1.355$ Mpc, and the formation redshift $z_{f}=1.052$ Gyr, which is equivalent to an age of $\sim8$ Gyr.

We follow the orbits of individual halos in the cluster, as shown in the examples in Figure~\ref{orbits}. This figure not only shows the typical trajectories of galaxies in phase-space as they fall (and eventually settle) into the cluster, but also reveals the distortions that can appear due to projection effects.   
For each halo, we measure the time-evolving ram-pressures, assuming the $\beta$-model presented in Section~\ref{sec:rps} (see Table~\ref{Cluster_param} in particular).  We assume the halos contain a MW-like disk with the same properties listed in Table~\ref{MW_values}, and 
flag them as ``stripped" (i.e. non-detected in HI) once they have been truncated down to $r_t$ and have thus fallen below the detection limit of our survey.

\begin{figure}
\centering
\includegraphics[scale=0.46]{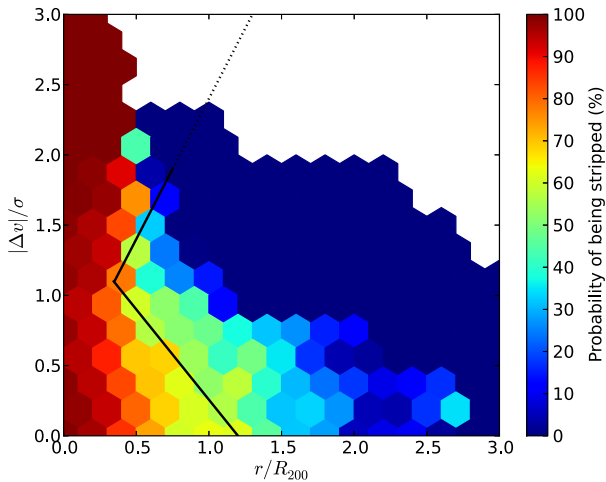}
\caption{Distribution in PPS of the probabilities that a MW-type galaxy has been stripped enough to be undetected by BUDHIES in a cluster with the ICM model of A963\_1. The red-orange areas trace the regions where galaxies have most likely been stripped of their gas. The thick black line is the same as in Figure~\ref{PPS_obs1}, that shows the boundary between the region where most HI-detected galaxies are (right of the line) and the region without HI-detections (left of the line). \label{marioplot}}
\end{figure}

In the top panels of Figure~\ref{simsplot}, we plot the final positions (at z=0) in a phase-space diagram, using only the halos that are infalling into the cluster for the first time, and distinguish the `HI detected' (open light green circles) from the `non-detected' or stripped galaxies (filled dark green circles).  It is noticeable how galaxies drop out of our detection limit as they pass through the ram-pressure stripping cone described in Section~\ref{sec:rps} (area left of the green dashed line in the right-hand panels of Figure~\ref{simsplot}). This effect is visible in both 3D and projected phase-space, although it is less distinct in projection. In the middle panels of the figure we plot all halos (infalling and virialized), and see that the non-detected galaxies are  not only in this cone, but elongate along a ``wedge", as indicated by the 
dashed-black line in PPS. This line is not an exact boundary, but a rough indication of the region where galaxies have a higher probability of having been in the cluster for more than a pericentric passage. 


From Figure~\ref{simsplot} we can see how the position of the simulated galaxies in PPS evolve with cosmic time. On first infall, the galaxies distribute along the escape-velocity caustic we saw in Figure~\ref{PPS_obs1}, avoiding low $r$ and low $v$ values (top panels of Figure~\ref{simsplot}). As the cluster evolves and becomes virialized, the galaxies cluster in the low $r$ and low $v$ region, as shown in the middle panels of the Figure (filled darker symbols, see also example orbits in Figure~\ref{orbits}). This shows the capability of PPS to predict, in a statistical sense, the time since infall of cluster galaxies. 

In order to ensure that dynamical friction is of the correct magnitude in our results, we subsample our halos, creating a `MW-like' subsample with initial halo mass: $0.25 \times 10^{11} - 7.5 \times 10^{12} M_{\odot}$. This approximately selects halos with the correct mass to match the stellar mass of our MW-like model \citep[see Eq. 3 in][]{Guo2010}. There are 96 in total in this mass range (bottom panel of Figure~\ref{simsplot}). 
By subsampling our halos, we are also assuring that there is a good match between the disk mass and the halos, although the general trends observed in the middle and bottom panels of Figure~\ref{simsplot} are not dissimilar. To be safe, however, we only use the sample plotted in the bottom panels in our analysis. 

To project the 3D phase-space, we randomize the line-of-sight to the cluster, spinning the cluster by a random angle about the x-axis, then another random angle about the y-axis. The line-of-sight is then chosen down the z-axis, with $r=\sqrt{x^2+y^2}$, and $v_{los}=v_z$. 

Then, we compute the probability, per PPS bin, that a galaxy has been stripped. To do this, we spin the cluster 1000 times and build up an average in each pixel of the PPS diagram (fraction of HI non-detected to the total number that lie in that pixel). This is shown in Figure~\ref{marioplot}, where red-orange regions trace the highest probabilities for a galaxy to have been stripped. In the figure, we have overplotted the line separating HI-detections from non-detections in the observed PPS shown in Figure~\ref{PPS_obs1}. The model predictions for where the effect of ram-pressure is strongest in PPS is in striking agreement with our observations, qualitatively reproducing the region in PPS where there are no HI-detected galaxies (left of the line). Although the results shown in Figure~\ref{marioplot} were computed specifically for MW-type galaxies in a massive cluster (like A963\_1), and assumed the HI detection threshold of BUDHIES, this PPS probability map can be used in other cluster studies as a guideline for where ram-pressure will be at play, provided the galaxy and cluster properties are not dramatically dissimilar from this study (see Appendix \ref{ap:rpmodel}). 
The virialized ``wedge'' will not change significantly. What will mostly change is the location of the ram-pressure stripping cone, that depends on the truncation radius (or HI detection limit).

\subsection{The effect of ram pressure in HI stripping}
\label{sec:bigpic}

\begin{figure*}
\centering
\includegraphics[scale=0.76]{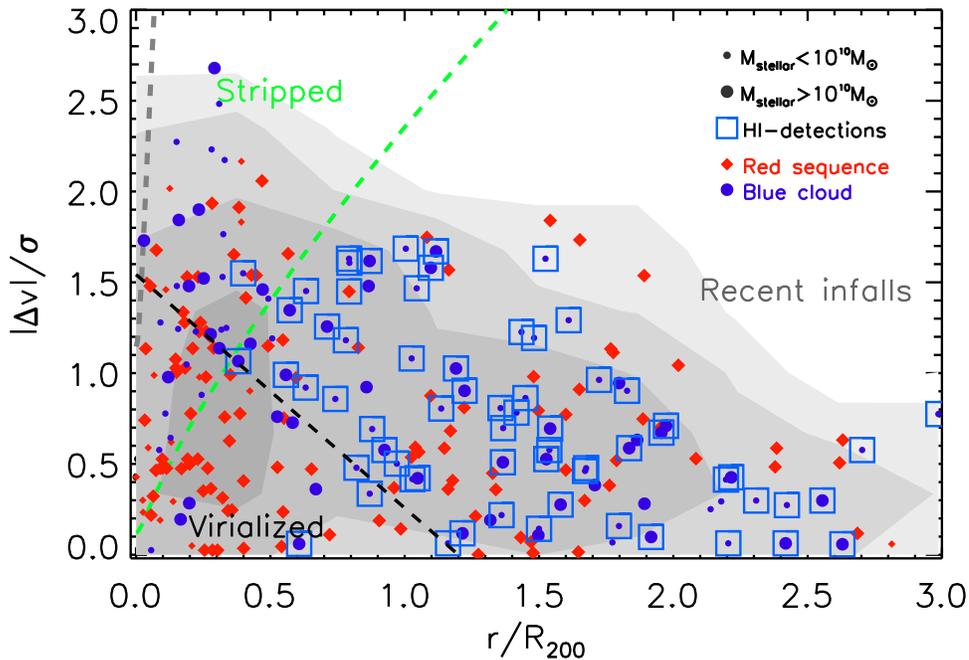}
\caption{The same PPS as in Figure~\ref{PPS_obs} is shown, but with added information:
As before, galaxies in the mass-limited sample ($M_{\star}>10^{10}M_{\odot}$) are represented with larger filled circles, HI-detected galaxies are enclosed by a blue open square, and grey countours follow the number density of the galaxies. In addition, galaxies are divided into red-sequence (red diamonds) and blue cloud (blue circles), as defined in the top panels of Figure~\ref{CMDs}. The dashed grey line delimits the area in PPS (left of the line) where a MW galaxy is expected to be completely stripped as it falls into the cluster. The dashed green line  delimits the area (to the left of the line) where the model galaxies are stripped enough to fall out of the detection limit of our survey (see Section~\ref{sec:sims}). As in previous figures, the additional dashed black line indicates the region in PPS where galaxies are most likely to have been in the cluster for more than a pericentric passage (i.e. the ``virialized" region). The region to the right of all dashed lines thus contains galaxies that have most likely recently joined the cluster. 
\label{PPS_obs}}
\end{figure*}


In this section, we combine the results of the previous sections, to 
link the effect of ram-pressure stripping with the orbital histories of galaxies in the studied cluster.

Figure~\ref{PPS_obs} shows the PPS of galaxies in A963\_1. 
As in Figure~\ref{PPS_obs1}, galaxies of different masses are shown with different symbol sizes, where the larger symbols correspond to the mass-limited sample. HI-detections 
are highlighted as before. We also distinguish between red-sequence and blue-cloud galaxies  
(defined in Figure~\ref{CMDs}).  
As in previous figures, the area to the left of the dashed green line indicates the ram-pressure stripping cone where galaxies will fall below our detection limit (see Section~\ref{sec:rps}), while the grey dashed line delimits the region of complete gas stripping. The area below the black dashed line instead, represents the ``virialized" zone, and the area outside the stripping and ``virialized'' regions thus contains galaxies that have recently joined the cluster, as labeled. 
In Table~\ref{tab_frac}, we further list the fraction of different galaxy populations (for the mass limited sample only) in all the regions of PPS defined. 

The most striking result of Figure~\ref{PPS_obs} is that both the stripping and virialized areas are almost completely devoid of HI-detections, and that their combined areas remarkably coincide with the region where simulated galaxies that have been stripped of their gas are located in PPS (left of line in Figure~\ref{marioplot}), regardless of their time since infall. 
%
%
%
Note that 
the stripped and ``virialized'' regions are not only devoid of HI-detections, but are also dominated by red galaxies (especially when considering the ``virialized'' region without the overlapping stripped cone, see Table~\ref{tab_frac}). This is consistent with galaxies aging as they spend more time in the cluster. 
The ``Recent infall'' region on the other hand is full of blue galaxies, likely coming from the field, as well as a significant fraction of red cluster members.  

Taking the mass-limited sample, we find that only a minority of the few blue galaxies in the ``stripped" and ``virialized" regions (or ``HI-poor" region) are detected in HI (9\%). 
Inversely, outside these regions (in the ``recent infalls" zone), we see that the majority of the blue cluster galaxies are detected in HI (67\%). 
Our results strongly suggest that field galaxies (typically blue, gas rich, star-forming spirals) that enter a massive cluster, will experience HI gas stripping on their first infall (i.e. first passage through the ram-pressure stripping cone), and that the intensity of the ram-pressure they experience is sufficient to make them 
undetectable by BUDHIES. 

\begin{figure}
\centering
\includegraphics[scale=0.45]{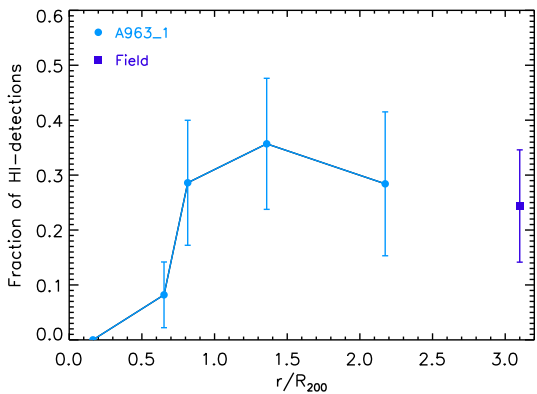}
\caption{Fraction of HI-detected 
galaxies in A963\_1 as a function of $r/R_{200}$ (solid blue line). The blue square at (arbitrarily) large radii corresponds to the field fraction in BUDHIES, for comparison. \label{rad_trends}}
\end{figure}

Although our galaxy sample is very incomplete at $M_{\star}<10^{10}M_{\odot}$ it is interesting to note that the ``HI-poor" region seems to be the same for low and high mass galaxies. This is consistent with the results shown in Appendix~\ref{ap:rpmodel}, where we found that variations in the mass of the model galaxy does not change significantly the predicted regions in PPS where ram-pressure stripping is most effective (except for very massive galaxies that are harder to strip). These two findings thus suggest that the orbital histories of the galaxies in the cluster are more important than their mass, to determine the amount of stripping they have experienced.  

As mentioned above, there is a noticeable fraction of red (and HI-poor) galaxies in the ``recent infall" region of PPS (59\%, see Table~\ref{tab_frac}), as well as some blue galaxies that are not detected in HI (33\%). 
The presence of stripped galaxies in the cluster outskirts is not a new finding. In the Virgo cluster, strongly stripped HI disks have been observed in the outskirts \citep{CrowlKenney2008}. Their morphologies and stellar populations, together with an observational estimate of their orbital histories, suggest these galaxies have been stripped in the outskirts of clusters, before crossing the centre \citep{Chung2009,CrowlKenney2008,Kenney2004}.


Using cosmological simulations of cluster formation and evolution, \citet{Tonnesen2007}  found that although  gas stripping occurs primarily in the central region of clusters, it is an important mechanism out to the virial radius. This is due to the wide scatter in ram-pressure strength that galaxies can experience at fixed radius. They also point out that the timescale for complete gas removal is $\geqslant1$ Gyr, and that galaxies in the cluster periphery ($r>2.4$ Mpc) often accrete cool gas. More recent simulations by \citet{Cen2014} further highlight the fact that the cluster environment can have an impact on the gas content of galaxies out to large radii, and that galaxies lose their cold gas as they fall into clusters, within a single radial round-trip. 
It is still unclear, however, whether the observational evidence for ram-pressure stripping outside clusters is related to the non-smooth density distribution of the ICM.

An alternative explanation for the presence of quenched galaxies\footnote{By ``quenched galaxies", we are referring to those galaxies with quenched star-formation} in the cluster outskirts (in PPS) is that not all of the galaxies in the outskirts of the cluster are necessarily falling in for the first time, but backsplashing instead. 
The amplitude of the backsplash after first infall and the time they will take ``virializing" depend on a number of factors, which makes the identification of backsplash galaxies in PPS very difficult. In fact, 
the regions where these galaxies tend to be found also host other galaxy populations, such as galaxies on their first infall or older cluster members \citep[e.g.][]{Oman2013}. This means that backsplashing is not a dominant effect at any particular PPS location. In particular, backsplashing is expected to be insignificant at large radii \citep[$r\gtrsim2 \times R_{200}$, see e.g.][]{Bahe2013,Muriel2014}, so it cannot account for all of the observed quenched fraction in the ``Recent infall'' region. Moreover, our own simulations, did not show significant quenching in this area of PPS (Figure~\ref{marioplot}), when we applied the ram-pressure stripping prescription to it.  This implies that not all red galaxies outside the ``virialized'' region can be attributed to backsplashing. An additional effect is needed to account for the high fraction of red galaxies out to the largest $r$. One likely possibility is that some galaxies had already been quenched by the time they fell into the cluster (e.g. though group pre-processing).  This idea is supported by the findings of \citet{Haines2007}, who showed (using SDSS) that the fraction of passively-evolving galaxies in the field can be as high as 50\%, when considering the most luminous systems. It is thus possible that some of these red galaxies in the outskirts of A963\_1 were already red when accreted into the cluster from the field or a lower-density environment. 
In fact, it is possible that a portion of these galaxies could have been  ``pre-processed'' (e.g. via mergers) in smaller groups that were accreted into the cluster. This idea is supported by our previous study of the distribution of HI-detected galaxies in the forming cluster A2192\_1 \citep{Jaffe2012}. We will investigate this further in future papers by studying the HI-content of galaxies in the small infalling groups visible in X-rays (see Section~\ref{subsec:cluster}). 

\begin{figure*}
\centering
\includegraphics[scale=0.76]{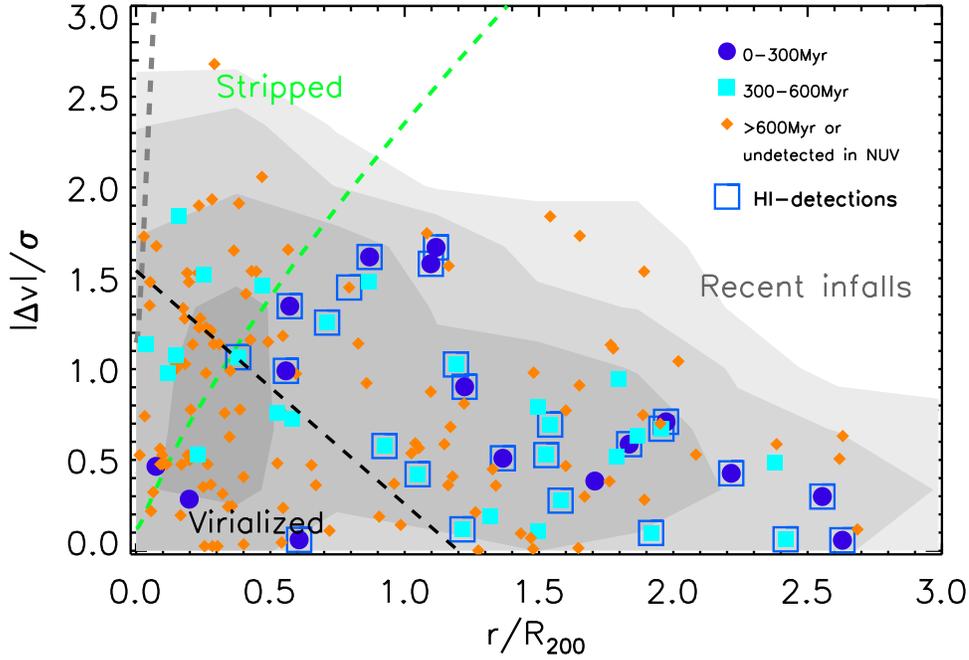}
\caption{
The same PPS of Figure~\ref{PPS_obs} is shown, but only for massive galaxies ($M_{\star}>10^{10} M_{\odot}$), this time colour-coded (and differentiated by symbols) by the stellar-population ages, inferred from their $NUV-r$ colours (see Section~\ref{subsec:cols}), as labeled. HI-detections are highlighted with blue open squares. 
Lines are as in Figure~\ref{PPS_obs}. \label{PPS_obs_UV}}
\end{figure*}

We also compare our results with those found by \citet{Bahe2013}, who recently analyzed the (hot and cold) gas content and star-formation rate of galaxies as a function of environment with hydrodynamic cosmological simulations. They find a significant decrease in the HI content and star formation with (3D) distance from cluster centre ($r_{_{3D}}$) out to very large distances ($r_{_{3D}}\sim5 \times R_{200}$, which is $\sim3\times R_{200}$ in projection, with scatter). Leaving aside the phase-space view for a moment, and only concentrating on $r/R_{200}$, we get the trends shown in Figure~\ref{rad_trends}, that can be compared with those found by \citet{Bahe2013}. The blue line corresponds to the fraction of HI-detected galaxies   corrected for HI incompleteness caused by the WSRT's primary beam attenuation. The correction was computed using the HI mass function of \citet{Martin2010} as reference. We note however that there might be a variation between  \citet{Martin2010}'s field HI mass function at low redshift, and that of A963\_1. We ignore this effect when calculating the corrections for Figure~\ref{rad_trends}, as we expect the difference to be within the plotted Poisson errors. An in-depth assessment of the HI mass function in BUDHIES is beyond the scope of this paper. We plan, however, to investigate the HI mass function as a function of both redshift and environment in a future paper.  

Despite the mentioned caveat, Figure~\ref{rad_trends} shows that the fraction of HI-detected galaxies in the cluster clearly  increases with $r/R_{200}$, until it reaches a plateau that matches our measurement for the field, within the errors. 
The general trend seen in BUDHIES is in agreement with those predicted by \citet{Bahe2013} (the red line in their Figure~2 corresponds to a cluster mass similar to that of A963\_1), when considering that their $r_{_{3D}}$ range ($0<r_{_{3D}}<5 \times R_{200}$) shrinks to our plotted range in projection. Although our HI fraction reaches the field value at smaller $r$, we note that the comparison can only be done qualitatively, as the absolute values of the fractions depend on their specific recipes and assumed thresholds. Measurements of HI-deficiencies in cluster galaxies at low z also show a significant  environmental effect only out to $\sim 2 R_{200}$ \citep{Solanes2001}, in agreement with our results. Our results however are not necessarily in disagreement with \citet{Bahe2013}'s conclusions, that evoke a combination of ram-pressure stripping, backsplashing and group-preprocessing to explain the observed trends.

\subsection{Star-formation quenching post HI stripping} 
\label{sec:Poststrip}

Even if field galaxies are affected by ram-pressure stripping as they first fall into the cluster, as proposed in ~\ref{sec:rps}, an infalling galaxy that crosses the BUDHIES ram-pressure cone (green dashed line) could continue to form stars in the remaining gas disk. Only when the galaxy has been completely stripped by the ICM (if it crosses the dashed grey line, passing very near the core of the ICM, or if it crosses less dense regions of the cluster sufficient times), will its star formation rate rapidly decline,  moving the galaxy from the blue cloud to the red sequence. At the same time, the galaxy will move in PPS until it settles into the virialized area. This scenario is supported by the properties of the blue galaxies inside the stripping cone, that have redder (although still blue) optical colours and on average, an earlier-type morphology (from the few available morphologies in the sample) with respect to the blue galaxies in the ``recent infall'' region. Moreover, when stacking the spectra of the few blue galaxies in the stripped zone for which we have AF2/WYFFOS data, we see  clear signs of $H_{\delta}$ absorption, which suggests declining star-formation activity.

\begin{table*}
\begin{tabular}{lcccccccc}
\hline
 	& Stripping cone & Virialized & Stripping$+$Virialized  & Stripping$-$Virialized & Virialized$-$Stripping& Recent infalls & Figure\\
 	 &  & & (\textbf{``HI-poor'' zone})  &  & & (\textbf{``HI-rich'' zone}) &\\ 	
\hline
$f_{r}$ &	79	& 83	& \textbf{80}	&72	&80	&\textbf{59}	& \ref{PPS_obs}	\\
$f_{b}$ &	21	& 17	& \textbf{20}	&28	&20	&\textbf{41}	&  \ref{PPS_obs}	\\
$f_{b,HI}$ &	0	& 14	& \textbf{9}	&0	&20	&\textbf{67}	&  \ref{PPS_obs}	\\
\hline
$f_{UV_{blue}}$&  4     & 7 & \textbf{5}	&6	&6	&\textbf{15}	&  \ref{PPS_obs_UV}	\\
$f_{UV_{green}}$& 14    & 11& \textbf{11}	&12	&9	&\textbf{25}	&  \ref{PPS_obs_UV}	\\
$f_{UV_{red}}$&   82    & 82& \textbf{84}	&88	&85	&\textbf{60}	&  \ref{PPS_obs_UV}	\\
\hline
\end{tabular}
\caption{Fraction (in \%) of different galaxy populations in the different regions of PPS. Following the column order: (1) the ram-pressure stripping cone (left of green dashed line in figures), (2) the virialized region (below black dashed line), (3) the combination of both of these regions (which coincides with the ``HI-poor" region of Figure~\ref{PPS_obs1}), (4) the ram-pressure stripping cone region excluding the overlapping area of the virialized zone, and (5) the region mostly populated with galaxies recently accreted to the clusters (right of dashed lines, which coincides with the ``HI-rich'' region of Figure~\ref{PPS_obs1}). 
$f_{r}$ and $f_{b}$ are the fractions of red and blue galaxies respectively, whilst $f_{b,HI}$ refers to the population of HI-detections within the blue galaxies. The last 3 rows show: $f_{UV_{blue}}$, the fraction of galaxies in the bluest NUV-r colour bin, or youngest stellar populations ($0-300$ Myr); $f_{UV_{green}}$, the intermediate-colour bin ($300-600$ Myr); and  $f_{UV_{red}}$, the reddest bin, that also contains galaxies that are not detected in NUV ($>600$ Myr), as defined in Figure~\ref{CMDs}.  
All fractions were corrected for spectroscopic completeness, considering the mass-limited sample only.}
\label{tab_frac}
\end{table*}
 
Turning  our attention 
to the star-formation in the stripped and non-stripped galaxies, 
we now analyze the location of UV-detected galaxies in PPS.  
In particular, it is very interesting to see where the bluer galaxies (in $NUV-r$)  are located in PPS with respect to the rest. 
This is shown in Figure~\ref{PPS_obs_UV} (large circles). 
Similarly to the HI-detected galaxies, the galaxies with the bluest $NUV-r$ colours (filled purple circles) tend to be located outside of the stripping cone, and are scarce in the ``virialized region'' (see Table~\ref{tab_frac}). This is consistent with the findings of Section~\ref{sec:bigpic}, since the blueness of these galaxies suggests that they host a very young stellar component ($\lesssim300$Myr), as discussed in Section~\ref{subsec:cols}. 
When we consider slightly redder galaxies (those with a stellar population component that is $\sim300-600$Myrs old, filled turquoise squares), we see that they occupy a larger area of PPS: although they are common outside the stripped area, they can also be found in the ram-pressure stripping cone, at the higher velocities, suggesting they have not been in the cluster for a long time. 
It is remarkable that the very-young and young galaxies are less common in the ``virialized'' area of PPS (see Figure~\ref{PPS_obs_UV} and Table~\ref{tab_frac}). 
Finally, if we consider the redder (older) galaxies (smaller symbols), it is clear that the distribution in PPS is less segregated, 
although they are predominant in the virialized region.   
The vast majority of these galaxies (including the ones not detected in NUV)  are red-sequence galaxies (see Figure~\ref{CMDs}), so it is not surprising that their distribution in PPS mimics that of red galaxies in Figure~\ref{PPS_obs}.  
As mentioned earlier, it is possible that some of these galaxies are indeed falling in to the cluster for the first time, but were quenched (or started to be quenched) in their previous environment. 

In summary, the distribution of UV-detections in PPS proved to be a good tool to study star-formation as a function of the galaxies' orbital histories. The results presented in this section, together with those presented in Section~\ref{sec:bigpic} support a scenario in which  blue, star-forming, late-type galaxies falling into the cluster from the field, are stripped of their HI gas (down to at least BUDHIES' detection limit) on their first passage through the ICM. After the gas is entirely removed, the star-formation is reduced and eventually quenched. This causes the ``virialized'' part of the cluster (in PPS) to be populated by red and ``dead'' systems. In addition, our analysis clearly shows that  not all galaxies take this evolutionary route. In fact, the presence of a significant fraction of gas-poor galaxies with old stellar populations in the infall region of the cluster requires an additional mechanism other than ram-pressure stripping (see discussion in Section~\ref{sec:bigpic}).  

\section{Summary and Conclusions}
\label{sec:conclu}

We use BUDHIES to investigate the role and timescale of HI gas stripping and later decline of the star-formation of galaxies in A963\_1, a massive regular cluster at $z=0.2$. We do this by constructing line-of-sight velocity vs. projected position phase-space diagrams of the cluster galaxies out to the cluster outskirts ($\sim 3 \times R_{200}$). Using cosmological simulations, these diagrams are  then used to infer (statistically) the orbital histories of the cluster galaxies.

Having a basic understanding of the cluster's assembly history allows to more effectively study the evolution of the cluster galaxies themselves, in particular, the transformation(s) they may experience as a consequence of having joined the harsh cluster environment. We investigate the effect of environment on the galaxies by studying the phase-space distribution of their HI, optical and NUV properties. 
The presence/absence of HI is an excellent indicator of environmental effects that lead to gas removal (e.g. ram pressure stripping), whilst the optical and UV colours can be used to distinguish galaxies with recent star-formation from those with old stellar populations. %

To interpret the observed segregation of HI-detected and non-detected galaxies in PPS, as well as the distribution of blue/red galaxies, we  
model the effect of ram-pressure stripping using the 
the prescription of \citet{GunnGott1972}, and adopting the HI mass limit of our survey ($\sim 2 \times 10^{9} M_{\odot}$), the ICM density profile of A963\_1, and the stellar masses of the observed cluster galaxies. 
We embed this model into the cosmological simulations so that we can study the effect of ram-pressure stripping as a function of the orbital histories of cluster galaxies. We then predict the regions in phase-space where galaxies are expected to be stripped of their gas. In particular, we create a map that shows, per phase-space bin, the probability that a galaxy has been sufficiently stripped to be undetected by our survey. 
The predicted map coincides strikingly well with the location of BUDHIES' HI-detected and non-detected blue (late-type) galaxies.

Our findings strongly suggest that HI gas in late-type galaxies can be stripped (down to at least the detection limit of our survey) during their first infall into the cluster.  In phase-space, the stripping will occur once they have approached the dense ICM core and/or gained enough velocity to cross the ``stripping" area in phase-space (small clustercentric distances and/or high line-of-sight velocities). 
After this, they will oscillate in phase-space (as they bounce back and forth in the cluster, gaining and losing speed) until they settle near the cluster centre due to dynamical friction. This long ($\gtrsim4$Gyrs) process    gives rise to the dense ``virialized" region observed in PPS. During this time, the galaxies' gas reservoirs are expected to be completely exhausted, which in turn causes their star formation rates to be quenched. 

To test this scenario, and to study the evolution of the stellar populations of galaxies after gas stripping, we combine the HI-observations with optical ($B-R$) and NUV ($NUV-r$) colours in our analysis. 
Although we find that galaxy colours also segregate somewhat in phase-space, the trends are less clear than when using HI. However, the ``virialized'' part of the cluster is dominated by red-sequence  galaxies,  as expected, while the cluster outskirts or infall region has a mixed population of galaxies.  
Moreover, we find that optically blue galaxies are less rare than HI-detected galaxies in the ``stripped'' region of phase-space. 
This is consistent with blue, star-forming galaxies first losing their gas reservoirs via ram-pressure stripping, and slowly declining their star-formation activity, as described above. In this scenario, the ``stripped''  galaxies (i.e. the ones not detected by BUDHIES) will still have blue colours for a while, until the gas is completely removed. Once this happens, the gas-poor galaxies will transit to the red sequence. %
The distribution of the galaxies' NUV colours in phase-space further support this scenario. In particular, we see how the bluest galaxies in $NUV-r$ (i.e. the ones with younger stellar populations) mimic the distribution of the HI-detected galaxies, while galaxies with ``green'' UV colours (intermediate-age stellar populations) are not only present in the cluster outskirts but are also found in the ``stripping'' region of phase-space. Finally, the galaxies with the oldest stellar populations (reddest NUV colours or undetected in NUV) dominate the ``virialized'' region of the cluster.

The described pathway to quench the star-formation in galaxies is, however, not unique. 
In fact, an additional insightful result from our analysis is the finding of 
a significant population of passive (red, gas-poor, early-type) galaxies in the outskirts or infall region of the cluster, that cannot be explained by our ram-pressure stripping model, or the effect of backsplashing. This finding suggests that  an additional mechanism must also be at play. 
One possibility is that galaxies are pre-processed in their previous enviroment (e.g. smaller groups, or the field) before they fall into clusters. We will investigate this in more detail in future papers. 
We conclude, however, that if a blue, gas-rich, star-forming galaxy has not been quenched by the time it falls into the cluster, it is most likely to experience gas removal as it first passes through the cluster's ICM due to the effect of ram-pressure.

Finally, this work  highlights the usefulness of  phase-space diagrams to infer  the orbital histories of galaxies in regular clusters,  and the effect of cluster processes such as ram-pressure stripping.  It should be noted, however, that this method cannot be robustly applied to very irregular (e.g. merging) clusters. 
Moreover, our phase-space analysis proved to be particularly insightful when using  sensitive tracers of environmental effects such as HI, in combination with optical and UV data.

\section*{Acknowledgments}

YJ gratefully acknowledges support by FONDECYT grant N. 3130476, and thanks Prof. Alfonso Arag\'on-Salamanca, Kyle Oman, and Jonathan Hernandez-Fernandez  for useful discussions on phase-space analysis. 
RS acknowledges FONDECYT grant N. 3120135, and  gratefully acknowledges Prof. Brad Gibson and Prof. Alexander Knebe for providing access to the cosmological simulations on which the modelling in this study  is based.  GC, and YKS gratefully acknowledge support by FONDECYT grants N. 3130480, and N. 3130470 respectively. 
Funding for the Sloan Digital Sky Survey (SDSS) has been provided by the Alfred P. Sloan Foundation, the Participating Institutions, the National Aeronautics and Space Administration, the National Science Foundation, the U.S. Department of Energy, the Japanese Monbukagakusho, and the Max Planck Society. The SDSS Web site is http://www.sdss.org/.
The SDSS is managed by the Astrophysical Research Consortium (ARC) for the Participating Institutions. The Participating Institutions are The University of Chicago, Fermilab, the Institute for Advanced Study, the Japan Participation Group, The Johns Hopkins University, Los Alamos National Laboratory, the Max-Planck-Institute for Astronomy (MPIA), the Max-Planck-Institute for Astrophysics (MPA), New Mexico State University, University of Pittsburgh, Princeton University, the United States Naval Observatory, and the University of Washington.






\appendix

\section{Sensitivity of the ram-pressure stripping model in PPS} 
\label{ap:rpmodel}

In this section we study the sensitivity of our ram-pressure stripping model (Section~\ref{sec:rps}) with variations of the cluster and galaxy model parameters. 

\begin{figure}
\centering
\includegraphics[scale=0.46]{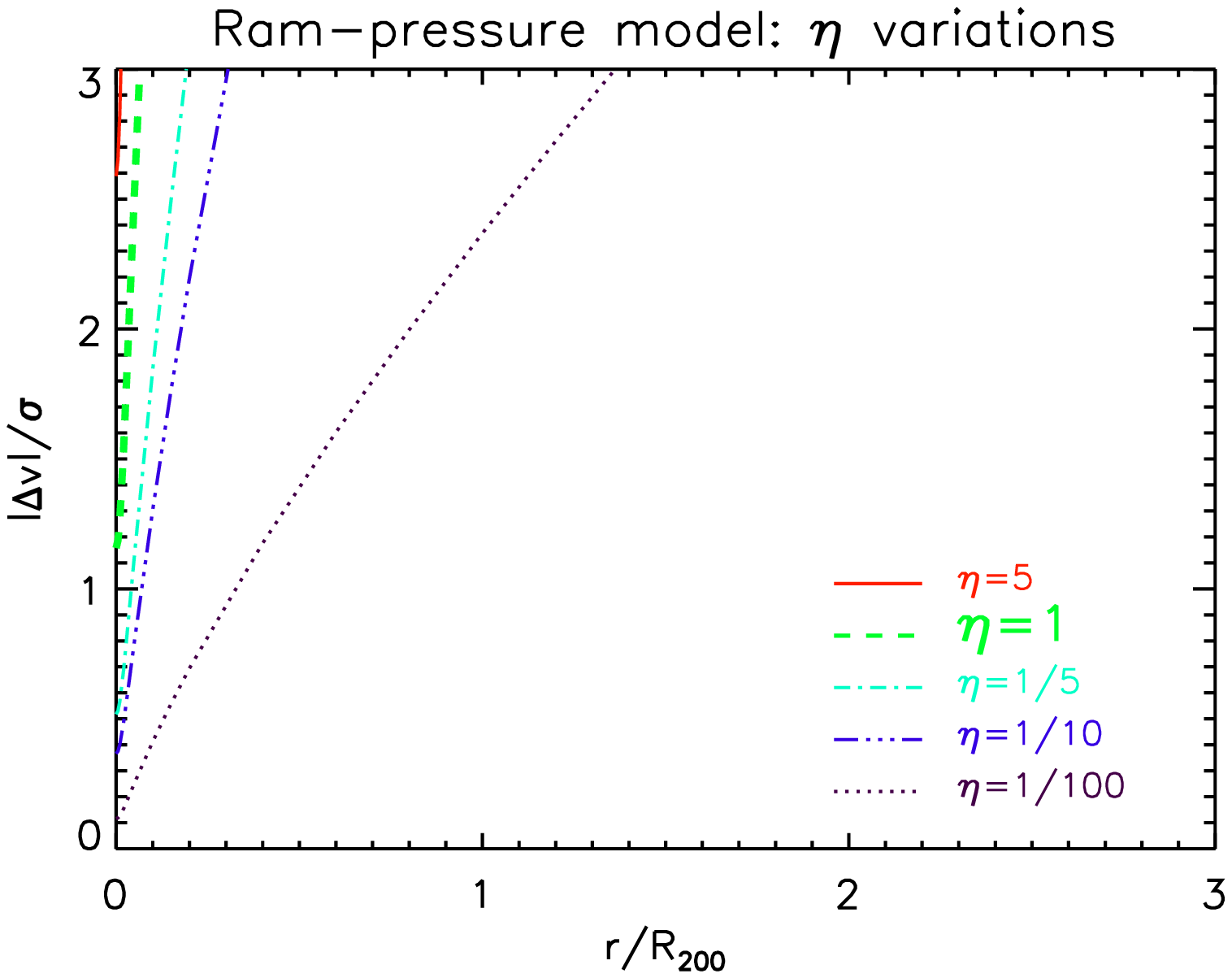}
\includegraphics[scale=0.46]{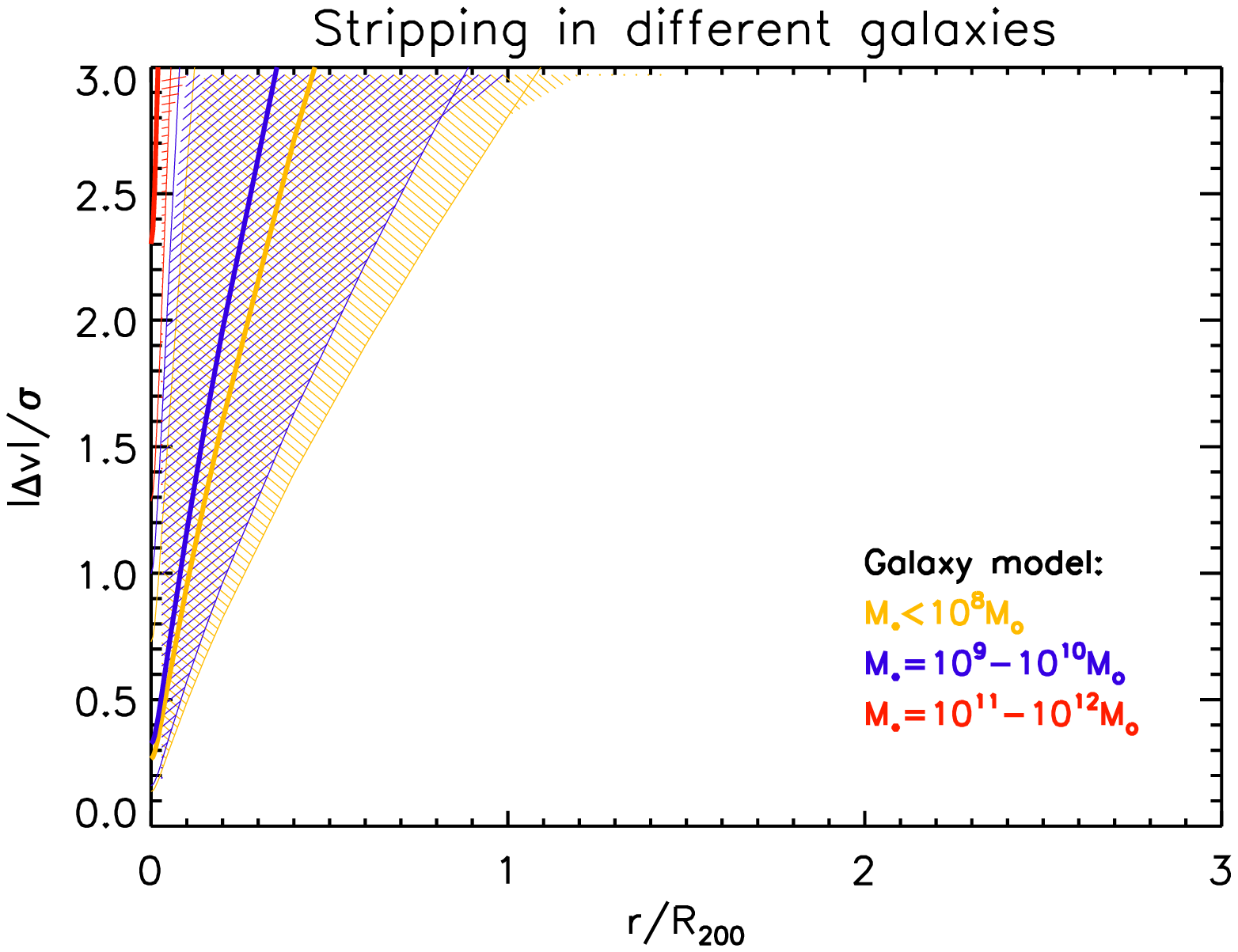}
\caption{\textit{Top:} The ram-pressure stripping model for a MW galaxy in A963\_1 is shown for different ram-pressure stripping intensities ($\eta=P_{ram}/\Pi_{gal}$), as labeled. \textit{Bottom:} The ram-pressure stripping model for cluster A963\_1 (using $\eta=1$) is tested against galaxies of different stellar mass, as labeled. The thick lines correspond to the model using the average scale length for that mass range (see text), and the shaded areas show the uncertainties introduced by the scatter in $R_{d}$. In both panels the axis range mimics that of Figure~\ref{PPS_obs} (left panel) for easy comparison.  \label{rps_variations}}
\end{figure}

We take A963\_1 and start by considering the case of a MW galaxy that is completely stripped: truncation radius $r^{\prime}=0$ and $\eta=1$, which corresponds to the dashed green line in the top panel of Figure~\ref{rps_variations} (same as dashed grey line in the left panel of Figure~\ref{PPS_obs}). This line  separates the region where $P_{ram}$ is able to strip all the gas in the modeled galaxy (left side of the line, $P_{ram}>\Pi_{gal}$) from the area where the anchoring force of the galaxy is still able to hold on to some of the gas  (right side, $P_{ram}<\Pi_{gal}$). 

Now, what happens if $P_{ram}$ is bigger or smaller than the value computed for A963\_1? Or when $\Pi_{gal}$ differs from that of the MW?  We illustrate this with the different lines in the top panel of Figure~\ref{rps_variations}. The lines correspond to different values of $\eta$, as indicated by the legend. For example, the blue line shows the region where $P_{ram}$ is 10 times higher than in A963\_1 (alternatively, it could also be the case of a galaxy with $\Pi_{gal}$ 10 times smaller than the MW, in a A963\_1 cluster). The red line instead corresponds to a cluster with ram-pressure 5 times weaker.  
This set of values attempts to account for the variation of $\Pi_{gal}$ and $\rho_{_{ICM}}$.  
As the figure shows there is a very small amount of variation, when considering variations of the order of 10 (blue and red lines). Significant deviations from the model only arise in very extreme cases (i.e. when $\eta$ is modified by a factor of 100). We thus conclude that the region where ram-pressure is expected to dominate (dashed greeen area in Figure~\ref{PPS_obs}) will not change significantly if the parameters used to model our cluster or our test galaxy are somewhat different.

Because not all of our galaxies are like the MW, we specifically test the sensitivity of our model (again, for A963\_1) with galaxies of different surface density, which depends on $R_{d}$, and $M_{\star}$, as shown by  equation~\ref{eq_gal_sd}. 

We take average values of $R_{d}$ in 3 bins of stellar mass from \citet[][see Figure 15]{Fathi2010} and compute $\Pi_{gal}$ in each case. The resulting ram-pressure stripping model (for $\eta=1$) is shown by the thick solid lines on the bottom panel of Figure~\ref{rps_variations}. Notice the stellar mass range considered is very large, from $<10^{8} M_{\odot}$, to $10^{11}-10^{12} M_{\odot}$. In this plot, our MW model ($M_{\star}=4.6\times 10^{10}M_{\odot}$) would lie somewhere between the blue and red line. We further accounted for the scatter or uncertainty in $R_{d}$ in each case, 
by computing the associated uncertainty in each model ($1\sigma$, shaded regions). The values used in each model are:
\begin{itemize}
\item For galaxies with $M_{\star}<10^{8} M_{\odot}$, the average scalelength is $\langle R_{d} \rangle=0.238\pm$0.094 kpc (green)
\item For $M_{\star}=10^{9} - 10^{10} M_{\odot}$, $\langle R_{d} \rangle=1.52 \pm 0.65$ kpc (blue)
\item For $M_{\star}=10^{11}-10^{12} M_{\odot}$,  $\langle R_{d} \rangle=5.73 \pm 1.94$ kpc (red). 
\end{itemize}
These values were taken directly from the text in Section~4.3 of \citet[][]{Fathi2010}. We did not use different mass bins to avoid interpolating $\langle R_{d} \rangle$ from their figure. 
The plot shows that for galaxies more massive than $10^{10} M_{\odot}$, and $R_{d}$ values within $1\sigma$ of those quoted in \citet[][]{Fathi2010}, the stripped region PPS is very similar.

We thus conclude that, there is little variability induced by different cluster models, and that for a mass-limited sample, the method is very robust in predicting the location in PPS where ram-pressure is expected to strip off the gas in galaxies.

\end{document}